# Guardian: Safe GPU Sharing in Multi-Tenant Environments

Manos Pavlidakis[*]
FORTH-ICS, Greece
manospavl@ics.forth.gr

Giorgos Vasiliadis[*,†]
Hellenic Mediterranean
University & FORTH-ICS, Greece
gvasil@ics.forth.gr

Anargyros Argyros[*,‡]
University of Crete & FORTH-ICS
Greece
argyrosan@ics.forth.gr

Stelios Mavridis[*]
FORTH-ICS, Greece
mavridis@ics.forth.gr

Antony Chazapis[*]
FORTH-ICS, Greece
chazapis@ics.forth.gr

Angelos Bilas[*,‡]
University of Crete & FORTH-ICS
Greece
bilas@ics.forth.gr

## ABSTRACT

Modern GPU applications, such as machine learning (ML), can only partially utilize GPUs, leading to GPU underutilization in cloud environments. Sharing GPUs across multiple applications from different tenants can improve resource utilization and consequently cost, energy, and power efficiency. However, GPU sharing creates memory safety concerns because kernels must share a single GPU address space. Existing spatial-sharing mechanisms either lack fault isolation for memory accesses or require static partitioning, which leads to limited deployability or low utilization.

In this paper, we present Guardian, a PTX-level bounds checking approach that provides memory isolation and supports dynamic GPU spatial-sharing. Guardian relies on three mechanisms: (1) It divides the common GPU address space into separate partitions for different applications. (2) It intercepts and checks all GPU related calls at the lowest level, fencing erroneous operations. (3) It instruments all GPU kernels at the PTX level –available in closed GPU libraries– fencing all kernel memory accesses outside application memory bounds. Guardian's approach is transparent to applications and supports real-life frameworks, such as Caffe and PyTorch, that issue billions of GPU kernels. Our evaluation shows that Guardian's overhead compared to native for such frameworks is between 4% - 12% and on average 9%.

## 1 INTRODUCTION

Graphic Processing Units (GPUs) have become necessary for accelerating applications, including machine learning (ML) and deep learning (DL) [17, 24, 51, 69]. Over the years, the compute and memory resources per GPU have been steadily growing to meet the rising demands for processing power [6, 7, 79, 80]. As a result of this trend towards more powerful GPUs, individual applications and, even more so, individual GPU kernels often fail to fully utilize the available resources, leading to GPU underutilization [6, 33, 43, 44, 47, 62, 63, 80, 82]. Sharing GPUs among applications of different users can improve resource utilization. GPUs today already support time-sharing, which allows switching from one application to another. Time-sharing involves costly context switches at the GPU level [33, 48, 77] and is not effective for applications that do not exhibit adequate parallelism [6, 47].

Several approaches propose GPU spatial-sharing [10, 18, 20, 23, 35, 52, 58, 66–68, 78, 82], enabling concurrent execution of applications to enhance GPU utilization. Spatial-sharing requires a single GPU context, i.e., a common GPU address space, and streams to execute kernels from different applications (and users) concurrently. However, this approach introduces a significant concern: applications in the same GPU address space can read or modify memory locations belonging to other applications [6, 12, 38, 49, 55]. NVIDIA's MPS [44] provides GPU spatial-sharing for applications started by the *same* user [44]. MPS offers memory protection but has serious fault isolation issues, i.e., when a kernel performs an out-of-bounds (OOB) access, it results in crashing any other co-running application [58, 71, 72, 84]. NVIDIA MIG [43] resolves these issues using a hardware mechanism that provides memory and fault isolation by statically partitioning the GPU at boot time and predefined sizes [33]. Each partition is assigned to a different user, but reconfiguring the GPU requires a hard reset resulting in high overhead [33, 58]. Previous work [33, 58] shows that this inflexible partitioning leads to underutilization. Other research [6] also proposes special hardware to enable protection, but practical use is limited due to this new hardware.

In this paper, we propose Guardian, a bounds checking software-based approach that provides transparent fault isolation from memory accesses while offering dynamic and flexible spatial GPU sharing. Guardian employs a client-server architecture that prevents applications from directly accessing the GPU, thereby eliminating the possibility of bypassing its protection mechanisms. To enforce memory protection and fault isolation, Guardian instruments the virtual assembly of GPU kernels offline. During runtime, it partitions GPU memory and assigns each partition to a specific application, ensuring that the instrumented kernels restrict every load and store operation to the application's designated memory space. Guardian effectively addresses three main challenges, as follows.

**Intercept GPU calls from closed-source libraries.** Guardian intercepts all GPU calls transparently at the CUDA runtime and driver library level by dynamically preloading its library into the application's execution. Previous approaches [14, 15, 18, 52, 61, 80] intercept all the CUDA stack, including the CUDA runtime, driver, and accelerated libraries. This is not sufficient for Guardian, though, because implicit CUDA calls, such as cudaMemcpy() and cudaLaunchKernel() (§7.7) performed from high-level CUDA library functions (e.g., cublasIsamax()) will go unprotected. Guardian forwards the intercepted GPU calls to the GPU manager (grdManager), which runs as a separate process and is the only entity with GPU access. This allows the grdManager to securely execute the GPU calls of different applications on a shared GPU.

**Fence illegal device accesses.** Application host memory is inherently protected because Guardian applications run as different processes. This is not the case for the device code, which runs on the same GPU context. To isolate the GPU address spaces of co-running applications, Guardian uses an allocator that divides the GPU memory into logical partitions assigned to applications. The allocation calls of each application are served from its partition, and every host-initiated transfer is checked at run-time to verify that it falls in a valid range. Guardian extracts and instruments the virtual assembly version of GPU kernels (PTX [45]), which are available even in closed-source libraries [64]. The PTX instrumentation includes the insertion of run-time checks to ensure that each pointer always falls in a valid range upon dereference.

**Lightweight bounds checking.** Address checking is a popular method for memory bounds protection, but it is costly because of the metadata management and the run-time checks. Reading the bounds from memory and inspecting if the pointer falls within these bounds incurs significant overhead [27]. To overcome this cost, Guardian follows a twofold approach. *First*, it uses contiguous partitions for each application, different from previous works [82]. This eliminates the need to store metadata for each allocation; instead, Guardian keeps only the start offset and the size of each partition, which can be stored in registers to avoid excessive memory fetches. These registers can be reused without adding significant register pressure to the execution of the GPU kernel. *Second*, it aligns the partitions in power-of-two sizes. This allows Guardian to optimize math operations (i.e., modulo) using fast bitwise instructions. In fact, Guardian adds only two bitwise instructions per load and store to isolate the memory partitions of different applications.

We implement Guardian for NVIDIA GPUs and evaluate it with real-life ML applications of Caffe and PyTorch, that link with closed-source GPU libraries. For Caffe and PyTorch, which invoke billions of GPU kernels, Guardian address fencing has on average 9% overhead compared to CUDA runtime (native), that offers time-sharing, whereas address checking is 1.7× worse than native. Guardian (protected) spatial-sharing is 4.84% slower than MPS. At the same time, it improves the total execution time of co-located applications by 37% compared to time-sharing, which is the alternative protected sharing mechanism used by other systems [28, 75, 80]. Finally, Guardian imposes minimal increase in register usage, and thus the transfer of excess variables from fast registers to slower

| Approach | OOB Fault Isolation | Dynamic Res. Alloc. | No HW support | Spatial sharing |
|---|---|---|---|---|
| Time-sharing [71, 75, 76, 80] | ✓ | ✓ | ✓ | - |
| GPU Streams [18, 52, 58, 82] | - | ✓ | ✓ | ✓ |
| MPS [44] | - | ✓ | ✓ | ✓ |
| MIG [43] | ✓ | - * | - | ✓ |
| Guardian | ✓ | ✓ | ✓ | ✓ |

**Table 1: Comparing Guardian with state-of-the-art GPU sharing approaches.** *MIG requires static GPU resource allocation.*

local memory (i.e., register spilling [55]), occurs only in 0.9% of PyTorch kernels.

The main contributions of this paper are:

- We design, implement, and evaluate Guardian, a novel system that offers *transparent* memory isolation for applications executing concurrently on a GPU without the need for static partitioning. We demonstrate its effectiveness using a broad range of kernels and complex, real-life ML frameworks [24, 51] that extensively use closed-source GPU libraries.
- We present a mechanism to intercept all GPU-related calls only at the CUDA runtime and driver library level. This allows transparently tracing and monitoring any GPU application or closed-source GPU library.
- We evaluate different bounds checking mechanisms implemented at the PTX-level and conclude that address fencing with bitwise operations is highly efficient and practical for protected GPU sharing.

## 2 MOTIVATION AND BACKGROUND

### 2.1 The Foundation for GPU Memory Isolation

A CUDA context is similar to a CPU process. As such, each CUDA application creates its own context during the first CUDA runtime call [65]. The context contains all the information regarding the resources used by an application, such as GPU memory, streams, cores, page table, and the GPU kernels to be used [55, 64]. An application that executes in a context cannot access memory locations used by an application in a different context [6, 83]. GPUs allow different contexts to time-share resources using a context switching mechanism that swaps the state of a GPU context to the GPU DRAM so that another context can be swapped in and run. A GPU stream is a sequence of operations (such as kernel executions, memory transfers, etc.) that are executed in order on a GPU. However, operations in different streams of the same context can execute concurrently, which enables parallelism and better resource utilization.

### 2.2 Limitations of Existing Sharing Solutions

Table 1 compares Guardian with other sharing approaches in terms of out-of-bound fault isolation, dynamic resource allocation, special hardware requirements, and spatial sharing.

**Time-sharing** offers memory and out-of-bound (OOB) fault isolation since it allows only one context to be active in the GPU at any time; and has been extensively used by previous works [14, 28, 71, 75, 76, 80, 81]. The device driver is responsible for allocating and managing resources belonging to a context. Upon a context switch, its resources are freed, and the translation lookaside buffer (TLB) is invalidated. Consequently, application data are protected, but at the cost of reduced GPU utilization, mainly due to the extra overheads of context switching [6, 77, 82]. Guardian eliminates such expensive context switchings, by enabling protected *spatial* GPU sharing, where applications run in parallel on the same GPU.

**GPU Streams**. Existing spatial sharing approaches [10, 18, 20, 23, 35, 52, 58, 66–68, 82] create a single GPU context with multiple streams [68] within, to enable concurrent application execution. This approach provides dynamic GPU resource allocation but without *any* memory protection. Guardian builds on the same paradigm and uses a GPU manager (i.e., grdManager) with multiple streams to enable sharing. The key distinction of Guardian is its bound-checking mechanisms, which ensure GPU memory protection and fault isolation, while the approaches mentioned above primarily focus on scheduling.

**NVIDIA's Multi-Process Service (MPS)** offers spatial sharing using the MPS server, which runs in the GPU and executes the kernels of different MPS clients or applications concurrently. MPS provides memory isolation due to the use of ASID TLBs [54] per MPS client. In particular, an access to an illegal address triggers a fault. However, when an MPS client fails, it leaves the MPS server and other MPS clients in an undefined state and may result in process hangs, corruptions, or failures [44, 58, 71, 72, 84]. We have found that when a kernel of an MPS client performs an illegal memory access, both the MPS server and other co-running clients are terminated. This is mainly because the MPS server allocates one copy of GPU storage and scheduling resources shared by all its clients [44] to enable sharing. As a result, MPS cannot be used in multi-tenant environments but only from cooperative applications [44]. Furthermore, MPS is closed-source and operates as a black box, prompting many research projects [23, 35, 52, 82] to develop their own open-source sharing mechanisms. These mechanisms are designed to host advanced scheduling policies and have demonstrated better performance compared to MPS [35, 52].

Guardian performs memory and fault isolation by preventing illegal accesses with bound checks, thus it is an effective solution for multi-tenant environments. In addition, the memory usage of Guardian remains constant and significantly lower than MPS due to the smaller amount of contexts created. In particular, Guardian creates only one context overall, while MPS creates a separate context per client. With just four clients (no data included) the GPU memory consumption of MPS (734MB) is 4× larger than Guardian (176MB), whereas with 16 clients it rises to 16× more (2.8GB vs. 176MB).

**NVIDIA's Multi-Instance GPU (MIG)** partitions statically high-end GPUs in completely isolated parts. Besides the limited applicability, recent works [6, 33] showed that MIG static partitioning leads to under-utilization. Moreover, changing from one partition scheme requires resetting the whole GPU, which results in high overhead (i.e., 100s of milliseconds [33, 58]).

Guardian uses a more dynamic GPU sharing scheme, similar to MPS [44] but with OOB fault isolation. Guardian different from MIG can: (1) Allocate partitions of different sizes according to each application's memory requirements. (2) Be turned-off on demand, so standalone or safe applications (checked with static analysis [30]) incur no overhead. (3) Be applied to different GPU models or GPUs of different vendors.

## 2.3 GPU Compilation Workflow

CUDA applications consist of the host (.cpp) and the device source code (.cu). The host code is compiled with clang or gcc, while the device code is with nvcc [22, 73]. The device source code is converted to the compiler Intermediate Representation (IR) format, which is then compiled, via cicc, to Parallel Thread eXecution (PTX) [45] assembly. PTX is a virtual assembly specification supported by the NVIDIA toolchain in all NVIDIA GPU architectures, past and future [55]. If necessary, the CUDA driver compiles the PTX code at runtime using just-in-time compilation and sends it to the new target device for execution [64]. This allows the generation of forward-compatible optimized machine code that runs on the target device. Nvcc always embeds the PTX representation of the device code in the target applications or libraries. nvcc also generates machine code for specific GPU architectures (using the ptxas assembler) and embeds this binary code to the application executable in the form of cuBIN files [64]. The generated PTX code and the cuBIN files are merged in a fatBIN file.

## 3 THREAT MODEL

Our work considers memory safety across kernels from different applications that share a GPU spatially in the cloud or other shared environments. Guardian prohibits applications from different users to read or modify each other's data in the host or device memory. Within the realm of GPU security concerns, our primary emphasis is on OOB fault isolation, as it represents a prominent threat to spatial GPU sharing.

We consider all GPU kernels, whether provided by individual users or from GPU libraries, unsafe. As a result, any instruction that performs loads or stores from a base address fetched from a destination register is considered unsafe and should be protected.

Regarding control flow, direct branch instructions are safe because they jump to labels defined inside a PTX file. The assembler will report errors if the labels are absent from the PTX file or are incorrect. On the contrary, *indirect* branch instructions (brx.idx) are unsafe because they use a register to index a statically defined array of labels. The register employed for indexing cannot be validated at compile time, potentially leading to out-of-bounds accesses.

PTX is an open and fully documented format, thus, parsing and instrumenting the needed instructions is straightforward. Due to the existence of PTX even in CUDA closed-source libraries (e.g., cuBLAS) and frameworks for forward compatibility, Guardian achieves 100% security coverage.

Guardian protects the GPU global and local memory. Registers and shared memory can not be accessed from co-running kernels [12, 30, 82], so they are safe. Heap, constant, texture, and unified memories are rarely used in ML frameworks, so we leave their protection as future work.

Guardian does not verify the integrity of data. Consequently, security issues related to exploiting GPU resource contention or side-channels [77], denial-of-service [39], and physical access [65] are outside the scope of this paper.

## 4 GUARDIAN DESIGN

The goal of Guardian is to provide GPU memory isolation for applications of different users that share spatially a GPU. Spatial GPU

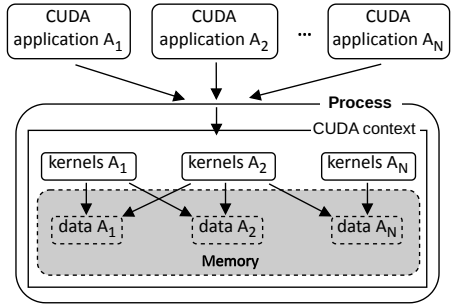

**Figure 1: Multi-tenant spatial GPU sharing, without Guardian. The common GPU context required for spatial-sharing allows applications to access each others memory.**

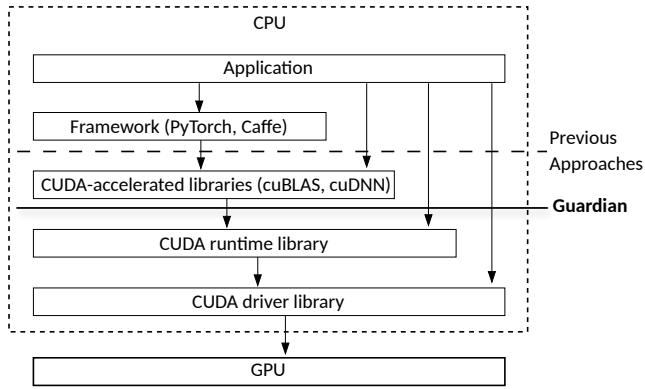

**Figure 2: Guardian CUDA library interception level versus previous approaches [14, 15, 18, 61]. Guardian intercepts only the CUDA runtime and driver APIs and *not* the level calls to CUDA accelerated libraries as in previous works.**

sharing requires a common CUDA context to execute kernels from different applications concurrently. Previous work [10, 18, 20, 23, 35, 52, 58, 66–68, 76, 78, 81] uses a separate process that creates that single context. Applications issue all their GPU tasks to this process, which enqueues to different streams, thus kernels can be executed concurrently. However, without proper protections, this approach allows GPU kernels to modify memory locations belonging to other applications, as shown in Figure 1.

Guardian is based on a client-server architecture, as shown in Figure 3, where the client is the CUDA application and the server is the grdManager responsible for the GPU access. Applications and the grdManager run in different address spaces; thus, we use an IPC channel and a separate shared memory segment per application to exchange CUDA calls and data similar to other CUDA interception approaches [18, 40, 52, 82]. Our approach uses three mechanisms shown in Figure 3 and described in more detail below. The dynamically loadable library (§4.1) intercepts CUDA calls and forwards them to a trusted process, the grdManager (§4.2) that executes GPU calls on behalf of the applications. The PTX-patcher (§4.3) applies bounds checking instructions (§4.4) to GPU kernels.

### 4.1 Dynamically Loadable Library

Guardian uses a dynamically linked library (i.e., grdLib) preloaded to the applications and transposes the default CUDA runtime and driver library, as shown in Figure 2. This library is generated automatically using the CUDA runtime and driver header files. These two CUDA libraries are the lowest interfaces for providing CUDA calls to manage GPU resources, such as allocating memory and launching kernels. CUDA applications and CUDA-accelerated libraries use the CUDA runtime interface and, to a lesser extent, the CUDA driver interface [14, 15]. The latter is used by applications only for specific, lower-level operations, such as explicit PTX (un)loading.

Guardian intercepts all GPU-related calls, including memory allocations, copies, and kernel launches. We ensure that all GPU calls are handled because the CUDA application links only with grdLib and not the native CUDA libraries, which have been removed from the LD_LIBRARY_PATH. Consequently, the application will fail to start if a CUDA call symbol does not exist in our library. Then, the intercepted CUDA calls are forwarded to another process, the grdManager (§4.2), which is the only entity with GPU access. This enables Guardian to securely perform any runtime checks necessary before executing the GPU operations on behalf of the applications.

The interception of CUDA calls is challenging for two reasons. First, the functions provided by CUDA-accelerated libraries invoke several *implicit* CUDA runtime calls, including memory allocations, transfers, and kernel launches. For example, a single cublasIsamax(), *implicitly* invokes several CUDA runtime calls, such as cudaMalloc(), cudaMemcpy(), and cudaLaunchKernel(), as shown in our evaluation (§ 7.7). Previous work [14, 15, 18, 52] treated such library calls as a black box, which is inadequate for Guardian because implicit CUDA calls can go unprotected.

Intercepting implicit calls requires applications to link with the *static* version of CUDA-accelerated libraries (e.g., libcublas_static.a) since only this version uses the shared version of CUDA runtime library (i.e., libcuda_rt.so). However, this is not always possible, as some applications are already built with the shared version of CUDA-accelerated libraries. To resolve this, we introduce a shim layer with two custom libraries (frontend and backend) to redirect shared library calls to their static versions, allowing interception without modifying the application build process. At runtime, the application runs using LD_PRELOAD to enforce them to use grdLib instead of native CUDA libraries. We also find that CUDA libraries dynamically load the CUDA driver library using dlopen() instead of linking with it. To prevent the original CUDA driver library from being loaded, we intercept dlopen() and provide the grdLib.

Second, CUDA libraries use an un-documented API function, namely cudaGetExportTable(), which returns tables of function pointers. The use of these functions depends on the application's need. For instance, we have found that large frameworks, such as PyTorch and Caffe, use about seven export tables containing more than 90 functions. By carefully rewriting these functions, Guardian can adequately intercept the CUDA runtime and driver libraries and run successfully real-world ML applications.

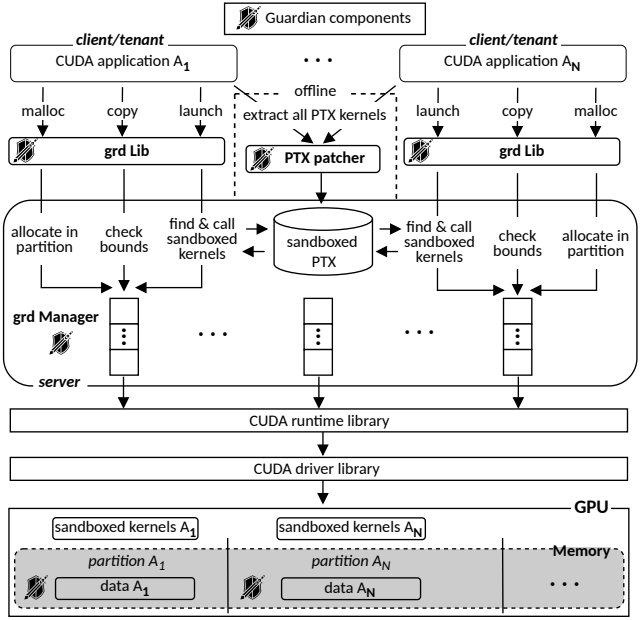

**Figure 3: Guardian online and offline (dashed annotated) mechanisms to allow protected spatial GPU sharing. Guardian intercepts the CUDA runtime interface used from applications and perform the necessary checks on memory allocations, transfers, and kernel executions. This allows kernels from different applications to execute concurrently on different memory partitions, eliminating illegal accesses.**

### 4.2 GPU Manager

*4.2.1 GPU Memory Partitioning.* GPU virtual memory is managed through the cudaMalloc()-family functions, which return arbitrary addresses upon each call. To perform GPU memory isolation across different applications, Guardian uses a custom allocator that initially reserves all GPU memory and splits it into partitions. Each partition is a contiguous memory block assigned exclusively to an application (or tenant-client). Contiguous memory partitions enable Guardian to offer isolation by checking that all memory accesses are always within the partition boundaries.

Guardian intercepts cudaMalloc() (malloc in Figure 3) and allocates memory in the application's partition. Similarly, our allocator marks the region in a partition as free by intercepting the cudaFree() function. For each application, we store the application id, the base address, and the partition size in a *partition bounds table* used at runtime. Guardian applications must specify their memory requirements at initialization, which is normal in cloud environments, where users buy instances with specific resources.

*4.2.2 Data Transfers.* Data transfers include operations that move data between the host and the GPU memory (e.g., cudaMemcpyH2D()) or within the GPU memory (e.g., cudaMemcpyD2D()). Even though these calls are initiated from the host, they refer to the same GPU address space; hence, applications can still perform memory operations to partitions of other applications. grdLib intercepts the memory management CUDA calls (copy in Figure 3) and sends them to the grdManager. The grdManager uses the *partition bounds table* to verify that the memory ranges are within the correct partition, and if this is true, it performs the transfer to the GPU. For all the cudaMemcpy-family functions, we check the destination and/or the source pointers.

*4.2.3 GPU Kernel Invocation.* The grdManager creates a new CUmodule for each PTX exported and patched during the offline phase (§4.3). A CUmodule is a CUDA code (PTX or cuBIN) unit that can be dynamically loaded and executed on the GPU. The CUmodules are then loaded into the current context using cuModuleLoadData(). A CUmodule can contain more than one kernel; hence, we use cuModuleGetFunction() to create a CUfunction handle for each kernel. The CUfunction handles are stored in a map, called *pointerToSymbol*, used to locate the appropriate kernel for execution.

At runtime, grdLib intercepts each kernel invocation via cudaLaunchKernel() and forwards this call to the grdManager, which then executes the corresponding sandboxed kernel, as shown in Figure 3. Every time a kernel is invoked for execution, Guardian performs a lookup at the *pointerToSymbol* table to find the CUfunction handle of the corresponding sandboxed kernel. Then, it adjusts the number of parameters accordingly; for address fencing (bitwise operation), it passes the mask and the base partition address (§4.3), whereas for address checking, the partition base and ending addresses. Each partition's information (base address, mask, or end address) is retrieved through the *partition bounds table*. Finally, the grdManager issues the sandboxed kernel using cuLaunchKernel(). When the grdManager detects that an application runs standalone, it issues a native kernel, avoiding the overhead implied by the extra instructions.

*4.2.4 Spatial Multiplexing.* To enable spatial-sharing, GPUs require a single context and CUDA streams provided by the GPU manager, i.e. grdManager, similar to previous works [10, 18, 20, 23, 35, 52, 58, 66–68, 76, 78, 81]. The CUDA runtime and driver call interception are performed on the application side using grdLib and are forwarded to the grdManager. As a result, applications do not create their own context; instead, they funnel their work to the GPU through the context of the grdManager. The grdManager is the only entity with access to the GPU and executes CUDA calls on behalf of applications after ensuring they are safe. Although we implement our grdManager, Guardian can be integrated into other systems [20, 33, 58] if their source code is available.

Guardian enforces all CUDA kernels and data transfers originating from a single application to be executed in-order from the grdManager. In contrast, kernels and data transfers from different applications will be executed concurrently using different streams. The difference between Guardian and previous works [18, 58] is the scheduling policy and not the sharing mechanism. Guardian implements a very basic scheduling policy that selects GPU calls from different applications in a round-robin fashion. Implementing more sophisticated scheduling policies is left as future work.

### 4.3 Offline Kernel Sandboxing

The PTX-patcher uses cuobjdump [41] to extract the PTX kernels from the application executable and the CUDA libraries (offline in Figure 3). The extracted PTX kernels are then sandboxed to ensure

```
1   .visible .entry kernel(
2   .param .u64 kernel_param_0,
3   .param .u32 kernel_param_1,
4   // Base address
5   .param .u64 kernel_base,
6   // Mask parameter
7   .param .u64 kernel_mask)
8   {
9     .reg .b32        %r<3>;
10    .reg .b64        %rd<5>;
11    ld.param.u64     %rd1, [kernel_param_0];
12    ld.param.u32     %r1, [kernel_param_1];
13
14    // Extra registers for base and mask
15    .reg .b64        %grdreg<3>;
16    // Load extra parameters to registers
17    ld.param.u64     %grdreg1, [kernel_base];
18    ld.param.u64     %grdreg2, [kernel_mask];
19
20    cvta.to.global.u64     %rd2, %rd1;
21    mov.u32          %r2, %tid.x;
22    mul.wide.s32     %rd3, %r1, 4;
23    add.s64          %rd4, %rd2, %rd3;
24
25    // Bit-wise And with mask
26    and.b64          %rd4, %rd4, %grdreg2;
27    // Bit-wise OR with base addr.
28    or.b64           %rd4, %rd4, %grdreg1;
29
30    st.global.u32    [%rd4], %r2;
31    ret;
32  }
```

**Listing 1: Sample sandboxed PTX CUDA kernel. Guardian address fencing (bitwise operations) implementation is explained with comments.**

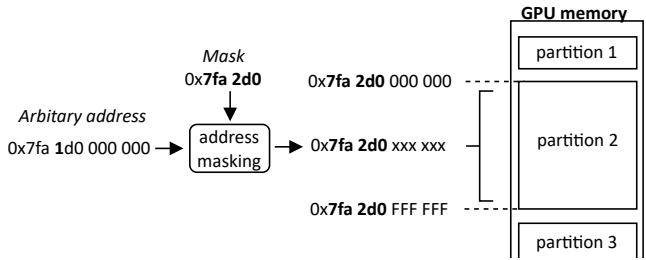

**Figure 4: Bitwise instructions mask addresses that fall outside a partition.**

they do not access data outside the correct partition boundaries. Listing 1 shows the sandboxed PTX code of a Kernel that assigns a value to each cell of an array ($A[tid] = j$). The original PTX (without sandboxing) consists of a kernel function definition that includes a list of parameters –lines 2 and 3. These parameters are addressable, read-only variables declared in the .param state space. Parameters are loaded to registers using ld.param instructions – lines 11 and 12. Each kernel allocates several registers used throughout the execution –lines 9 and 10. Then, the kernel uses these registers to load and store the values generated in each execution step –lines 20-23 and 30-31. The whole process is fully automated and covers the entire PTX ISA of CUDA 11.7.

Our patcher (1) adds two extra parameters in each kernel –lines 5 and 7, (2) defines two extra registers to load the mask and the base partition address parameter –line 15, (3) loads the extra parameters in the registers –lines 17-18, and (4) appends two bitwise instructions –lines 26 and 28– before every load/store. The bitwise AND operation is performed between the load/store address and the mask. The mask for each partition is calculated using the highest address and the partition size. For instance, if the partition starting address is 0x**7fa2d0**000000 and the partition size is 16 MB: the ending address is 0x**7fa2d0**FFFFFF and the mask is 0x**000000**FFFFFF (partition 2 in Figure 4). In any case, the number of zeros in the mask depends on the partition size. Then we use a bitwise OR between the address and the base address of a partition. The bitwise AND with the mask and the bitwise OR with the partition base address make an address outside the partition to start from the begging of the partition, i.e., wrap around, as shown in Figure 4. The illegal address that points to partition 1 (assigned to another application) due to the bitwise operations with the masking address, will finally point to partition 2. With this approach, only invalid or malicious kernels will wrap around and potentially corrupt their data. If memory corruption results to other execution issues (e.g., no convergence in ML applications) for invalid or malicious applications, the grdManager can utilize existing techniques [23, 53] to detect and terminate the endless kernel. Alternatively, Guardian can use address checking (§4.4) to detect invalid accesses and return from the kernel, but at a higher cost (§7.2).

Intel, AMD, and NVIDIA GPUs have two addressing modes for loading or storing data to memory [30, 45]. In the first case, the base address is loaded into the destination register (line 30 in Listing 1), while in the second, an offset is first added to the base address, and the result is loaded into the destination register (i.e., $ld.global$ %$val$, [%$base\_addr + offset$]). The same modes apply to stores and all off-chip memories. The PTX-patcher applies the bit-masking instructions directly to the base address for the first mode. For the second mode, the patcher calculates the new address by adding the offset to the base address and stores this in a new temporary register. Then, it applies masking instructions to this new address. Our patcher instruments .func in the same way as kernels (.entry). The .func directive denotes a function callable from both host and kernel code.

### 4.4 Bounds Checking Tradeoffs

Guardian currently supports three bounds-checking methods: One address checking and two address fencing approaches. Each approach has different requirements and can be dynamically utilized at runtime by Guardian to serve different purposes. First, address checking uses *conditional checks* to verify that the addresses used in load and store operations are in the correct partition. This approach detects out-of-bounds accesses and is more suitable for debugging purposes. Unlike address fencing, address checking can be used for partitions of arbitrary size but at a higher cost (80 cycles) because the Address Divergence Unit executes conditional checks.

Address fencing does not provide out-of-bounds detection, but it is more efficient and practical, making it sufficient for isolating concurrent applications. Address fencing with *modulo* applies the following instructions before every load and store: $fenced\_addr =$

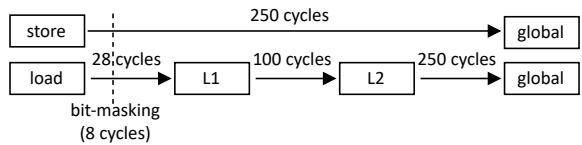

**Figure 5: Bit-masking latency (8-cycles) compared to latency of different memories.**

$partition\_base+((arbitary\_addr-partition\_base)\%partition\_size)$. CUDA ISA implements the 64-bit modulo operation via a function call that requires 2× more cycles than the 32-bit modulo implemented inline by NVIDIA. We implement the 64-bit modulo inline with three instructions and an extra parameter holding the $\frac{1}{parttition\_size}$. The extra parameter avoids the division's high overhead since it is also implemented via a function call. This approach implies less overhead (28 cycles) than address checking and still does not require partition alignment.

Address fencing with *bitwise* operations is the most lightweight compared to previous approaches because it requires 8 cycles –4 cycles per bitwise operation [3]. As shown in Figure 5, a load/store instruction requires 28 cycles if the data reside in L1-cache, whereas if the data is in global memory, it requires 220-350 cycles [7, 25]. In the rare case that all the data are in L1-cache (100% cache hit ratio), our approach implies 30% overhead, whereas in the typical case (data in global memory), all CUDA kernels add on average 3.5% (§7.4). Our approach requires the partitions to be in the power of two to cut down the extra instructions required to check a partition's upper and lower bounds. The power-of-two block size allocators restrict the number of concurrent applications, however PyTorch and TensorFlow use this type of allocator as default. Consequently, we choose to optimize the common case and leave the allocation issue as future work. We can use different allocator schemes in combination with the fencing/checking approach, e.g., to allocate GPU memory more progressively for each application. Such extensions, along with existing allocators [31, 37, 70, 74], could resolve the fragmentation of contiguous partitions and are very interesting to be examined as future work.

Guardian passes the mask and the base partition address to every kernel using two parameters. Using these parameters inside the kernel requires two extra registers. This does not lead to register spilling (§7.3) because GPU kernels use the minimum number [19, 55] of registers, and the nvcc compiler optimizes further register usage [34]. We also examined two other possible solutions: The first is a global map stored in GPU memory, but updating the map is prohibitively expensive. The second is to generate a different kernel binary for every partition with the mask hard-coded. This approach does not scale when multiple applications use thousands of kernels. The grdManager compiles at its initialization the sandboxed PTXs avoiding JIT overhead at runtime.

## 5 SECURITY ANALYSIS

In this section we evaluate the security properties of Guardian by describing possible attacks, and showing how our proposed design protects against them.

**CUDA kernel reuse attack.** PTX kernels according to Guardian are shared. However, Guardian executes only kernels that have been instrumented (1) with extra kernel parameters to pass the partition bounds at runtime and (2) bound checking instructions to secure illegal instructions. These steps guarantee that a GPU kernel, even if shared, always runs within the specific application's address space.

**Bypass Guardian checks.** With Guardian, applications do not have direct access to the GPU (no CUDA context per application) that could bypass our checks. To achieve that, Guardian intercepts and forwards all CUDA API calls at the runtime and driver level to a trusted component, the grdManager. grdManager is responsible for executing all GPU calls on behalf of applications and has exclusive access to the GPU.

**OOB fault isolation.** Guardian adds bounds checking instructions before every load/store and indirect branches. This way, Guardian ensures that a kernel cannot access memory outside its partition, leading to fault isolation.

**Buffer overflow.** Guardian does not protect or detect illegal accesses from a single application since it does not keep information per data buffer. Keeping information and checking each data buffer is costly regarding memory usage and performance. This is the case for debugging tools, as CUDA-MEMCHECK and cuCatch [60, 64].

## 6 EXPERIMENTAL METHODOLOGY

We now assess the performance of Guardian by answering the following questions:

(1) What is the impact of Guardian on GPU spatial-sharing compared to NVIDIA MPS (§7.1)?
(2) What is the overhead of Guardian on real-life applications – running standalone– compared to native execution and other protection approaches (§7.2)?
(3) What is the impact of address fencing (bitwise operation) on GPU register usage (§7.3)?
(4) What is the performance of address fencing (bitwise operation) at high cache hit ratios (§7.4) using different GPUs and access patterns (§7.5)?
(5) What is the cost of CUDA runtime and driver API interception (§7.6)?
(6) What are the implicit CUDA runtime and driver calls performed from closed-source libraries (§7.7)?

**Server platforms.** To evaluate Guardian we use two GPU models (Table 2), that are installed on two different servers. The first server is equipped with a Quadro RTX A4000 GPU, four AMD EPYC 7551P NUMA CPUs with 8 physical cores each (running at 3.0 GHz, hyper-threaded), and 128 GB of DRAM. To avoid passes over QPI/UPI, we pin applications to the cores closer to the GPU. The second server contains a GeForce RTX 3080 Ti, an Intel(R) Core i7-8700K CPU with 6 cores running at 3.70 GHz, and 32 GB of DRAM. Both servers have NVIDIA CUDA v11.7 with NVIDIA driver v.515 installed. All the experiments, except §7.5, are performed in the Quadro RTX A4000. Regarding GPU kernel scheduling, we use the default NVIDIA policy, namely leftover [21, 47].

**Applications and datasets.** To evaluate the overheads of Guardian under *real-world* scenarios, we use multiple neural networks from Caffe [24] and PyTorch [51] frameworks with large data sets that invoke billions of kernels and execute for hours and applications from the Rodinia benchmark suite [9]. Regarding ML applications, we run lenet, siamese, computer vision, and rnn neural networks with the mnist dataset [29], while for cifar10, the cifar dataset [26]. Both mnist and cifar datasets contain hundreds MBs of images. All the above neural networks are executed with 100 up to 500 epochs and invoke up to 142 million CUDA kernels. We also run experiments with imagenet dataset [56], which consists of 256 GB of images, using googlenet, alexnet, caffenet, vgg11, mobilenetv2, and resnet50 as neural networks. These networks invoke billions of kernels, and we run them for ten epochs leading to 99% accuracy. Table 3 shows the total number of kernels, functions, and the load/store (ld/st) instructions contained in the libraries and frameworks that we use in our evaluation. Guardian detects all these instructions and protects them. Regarding Rodinia, we increase the default dataset size by 10× and kernel execution time by 8×, compared to previous work, because the default values are small for executing on real systems.

| Specifications | RTX A4000 | RTX 3080 Ti |
|---|---|---|
| Compute Capability | 8.6 | 8.6 |
| #SMs | 48 | 80 |
| #CUDA cores | 6144 | 10240 |
| L1 (KB) | 128 | 128 |
| L2 (KB) | 4096 | 6144 |
| Global memory (GB) | 16 | 12 |
| #Registers / Thread | 255 | 255 |
| PCIe | v4 x16 | v4 x16 |
| L1 hit latency (cycles) | 28 [7, 25] | 28 [7, 25] |
| L2 hit latency (cycles) | 193 [7, 25] | 193 [7, 25] |
| Global memory BW (GB/s) | 448 | 912 |
| Error Correction Code | Yes | No |

Table 2: GPU specifications we use for the evaluation.

| Libraries/Frameworks | #kernels | #func | #total loads | #total stores |
|---|---|---|---|---|
| cuBlas (v11) | 4115 | 0 | 341249 | 106399 |
| cuFFT (v10) | 5173 | 4 | 175256 | 371932 |
| cuRAND (v10) | 204 | 0 | 4949 | 3610 |
| cuSPARSE (v11) | 4335 | 0 | 334694 | 101792 |
| Rodinia | 23 | 7 | 544 | 285 |
| Caffe | 1294 | 4 | 87267 | 32946 |
| PyTorch | 27987 | 319 | 2083978 | 857987 |

Table 3: Load and store instructions identified and safeguarded within the CUDA-accelerated libraries and frameworks used in our evaluation.

| Workloads with **same** apps | | | Workloads with **different** apps | | |
|---|---|---|---|---|---|
| ID | Name | Epochs per app | ID | Name | Epochs per app |
| A | 2xlenet | 500 | I | lenet-siamese | 500-50 |
| B | 4xlenet | 500 | J | siamese-cifar10 | 30-100 |
| C | 2xcifar10 | 100 | K | 2xlenet-siamese-2xcifar10 | 500-30-100 |
| D | 4xcifar10 | 100 | L | 3xlenet-siamese-2xcifar10 | 500-30-100 |
| E | 2xgaussian | - | M | hotspot-guassian | - |
| F | 4xgaussian | - | N | gaussian-lavamd | - |
| G | 2xlavamd | - | O | particle-hotspot | - |
| H | 4xlavamd | - | P | gaussian-hotspot-lavamd-particle | - |

Table 4: Mixes of workloads used for assessing the performance of Guardian under GPU sharing.

**Workloads.** From Caffe, PyTorch, and Rodinia, we create a set of workloads shown in Table 4, to evaluate Guardian under GPU sharing. Each workload is a mix of compute- and data-intensive applications and covers scenarios in which applications compete and stress the GPU resources. As in previous works [11, 33, 59], we create workloads with 2-6 concurrent clients. The workloads A-H use multiple instances of the same application, while I-P includes different applications. To ensure that application executions overlap, we appropriately modify the number of epochs of each application, affecting the total execution time. We also vary the batch size to increase memory usage in each application from 500 MB to 2 GBs. To assess the applicability and coverage of Guardian, we use CUDA library examples [42] for cuBLAS, cuFFT, and cuSPARSE libraries. These examples include more than 30 library calls that are not in the ML frameworks used.

**Performance measurements.** We use Nsight [32] to profile GPU kernel execution and collect metrics, such as cache hits, GPU call latencies, and kernel invocations. We use the Xptxas=-v compiler flag to measure register and constant memory used by Guardian's sandboxed kernels. We use the rdtsc instruction to measure host calls' duration (in CPU cycles).

**Baseline and Guardian Deployments.** Regarding *GPU sharing* we use four deployments. *Native* (baseline) uses the default CUDA runtime environment, which offers time-sharing with memory protection and fault isolation. The other three setups provide GPU spatial-sharing: NVIDIA MPS [44] allows concurrent execution of multiple kernels with memory protection without fault isolation. Guardian without protection is based on Arax [52] implementation of spatial-sharing. This setup intercepts GPU calls but performs neither checks nor instrumentation. Guardian with address fencing (bitwise operation) is our main approach for protection and fault isolation using bit-masking. Regarding G-NET [82] that uses network functions for its evaluation, we extrapolate its protection mechanism for ML applications using address checking. Mask [6] uses the Mosaic simulator [5] for its evaluation, and we omit to compare directly with Guardian. Finally, MIG [43] statically partitions high-end NVIDIA GPUs, leaving GPU resources underutilized, making a comparison less relevant.

We run *standalone neural networks* to isolate Guardian protection overheads. This is essential because spatial-sharing intensifies resource contention, which may increase the latency of GPU loads and stores, which may amortize our overheads. Regarding this scenario, we use: (a) *Native* CUDA as a baseline. (b) Guardian *without protection* this setup models the overhead of intercepting and forwarding CUDA calls to the grdManager. (c) Guardian with *address checking*, to evaluate control flow instructions. (d) Guardian with *address fencing modulo operation* to measure the overhead of our inline modulo instruction. (f) Guardian with *address fencing bitwise operation* to appraise bitwise instructions.

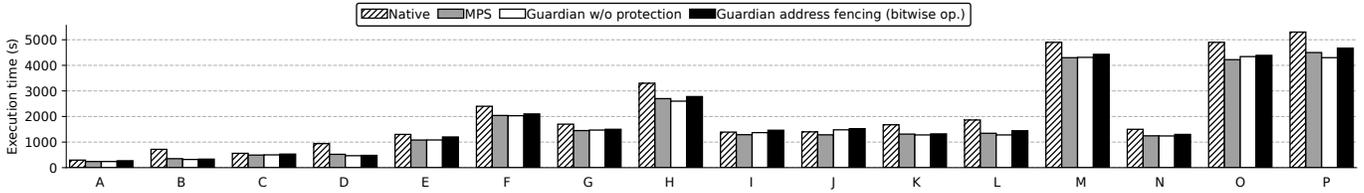

**Figure 6: Multi-tenant GPU sharing using native CUDA time-sharing, MPS spatial- sharing (no fault isolation), Guardian spatial-sharing (no fault isolation), and Guardian spatial-sharing with address fencing (bitwise operation).**

## 7 EXPERIMENTAL EVALUATION
### 7.1 Impact of Guardian at GPU Sharing

Figure 6 shows the execution times of *Native*, *MPS*, *Guardian without protection* and *Guardian address fencing bitwise operation* for the workloads of Table 4. Comparing Guardian address fencing (bitwise operation) to MPS, our approach is, on average, 4.84% slower due to the extra checks enforced to prevent out-of-bounds accesses. When we turn off these checks in Guardian (no protection), the execution times achieved are 0.05% worse than MPS. In high resource contention, as in workloads I-P, the overheads of Guardian address fencing are lower on average 3.2% since our overheads are amortized. Guardian without protection performs better than MPS in workload with thousands of pending kernels (D, P, K, H) because the MPS server becomes a bottleneck [52].

Finally, we compare spatial and temporal sharing, which is the default mechanism used from many previous works [8, 28, 75, 80] because it ensures protection. Guardian address fencing is, on average 23% faster than native, while in some cases, it is up to 2× faster due to parallel kernel execution. We note that the performance improvements of spatial sharing are primarily affected by the resources required from the concurrently executed workloads. In cases where the resources needed are low, as in workloads B and D, the benefits are more prominent, i.e., 2×, while the performance gap is reduced for more resource-intensive workloads.

### 7.2 Guardian Overheads Compared to Other Approaches

Figures 7 and 8 plot the times of the standalone execution for ML applications that use CUDA-accelerated libraries (Table 3), using *Native*, *Guardian without protection*, *Guardian address fencing (bitwise and modulo)*, and *Guardian address checking*. The training phase for lenet, siamese, cifar10 issues up to 142 million kernels, whereas googlenet, alexnet, caffenet, vgg11, mobilenetv2, and resnet50 issue billions. The inference phase issues up to 8 million kernels.

Figure 7(a) shows lenet, siamese, and cifar10 training, while Figure 7(b) shows the inference phase of the same neural networks. Guardian has between 5.9% up to 12% overhead compared to the unprotected native CUDA. The Guardian without protection approach includes the interception of CUDA calls and the search in the *pointerToSymbol* table to find the appropriate sandboxed kernel. The kernel issued in the GPU does not contain the bit-masking instructions, while transfer instructions do not contain the out-of-bounds checks added from Guardian address fencing. The Guardian without protection approach has an overhead from 3.7% to 10% compared to native. By comparing Guardian address fencing (bitwise operation) and Guardian without protection, the overhead of address fencing is between 1.05% up to 4.3%. Thus, the overhead of bounds checking (in transfers and kernels) is 2.9% on average.

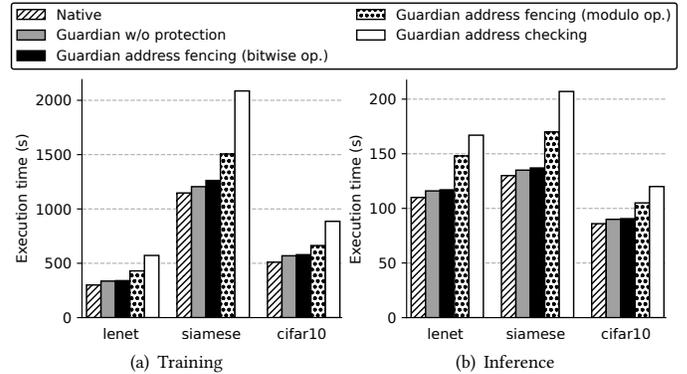

**Figure 7: Comparison of address fencing (bitwise) with other approaches, using Caffe with *mnist and cifar* dataset.**

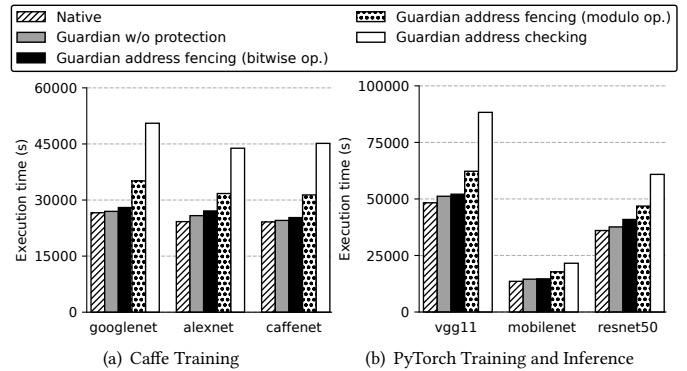

**Figure 8: Comparison of address fencing (bitwise) with other approaches, running Caffe and PyTorch with *imagenet*.**

Figure 8(a) shows googlenet, alexnet, and caffenet training. Guardian address fencing (bitwise operation) has between 4.5% up to

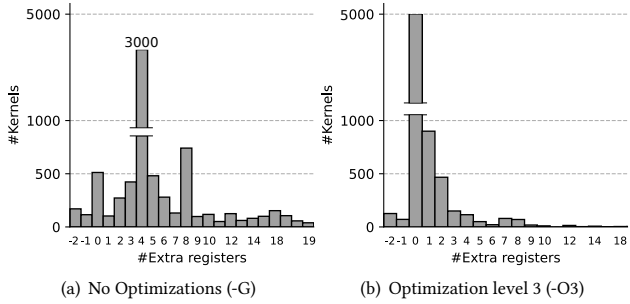

(a) No Optimizations (-G)  (b) Optimization level 3 (-O3)

Figure 9: Guardian's per thread register usage vs native.

10% overhead compared to the unprotected native CUDA. Guardian without protection, which includes only the interception of CUDA calls, has an overhead from 1.36% to 6% compared to native. By comparing Guardian address fencing (bitwise operation) and Guardian without protection, the overhead of Guardian is between 2.9% up to 4.3%. Figure 8(b) shows vgg11, mobilenet, and resnet50 training and inference using PyTorch. The overhead of Guardian for call interception is, on average 5.5% (native vs. Guardian without protection). The overhead of Guardian address fencing compared to Guardian without protection is on average 7.6%. Compared to existing CPU [1, 27] approaches, GPU bounds checking has lower overhead, since kernels have simpler access patterns.

Our optimized modulo –without function call– approach, namely address fencing modulo operation, increases the execution time by 29% on average compared to native, due to the addition of seven extra instructions. Conditional checks increase execution time by 1.7× on average compared to native. This is because the branch instructions are more expensive compared to bitwise operations. In the addressing mode that uses address+offset we add up to eight instructions (32cycles) to check the bounds of each partition.

### 7.3 Address Fencing Impact on Register Usage

Figure 9 shows the number of registers used for storing the address mask and the base address in the address fencing bitwise approach. Figure 9(a) shows the additional registers used from our approach when compiling the PTX without any optimization flag, whereas Figure 9(b) shows full optimizations. The lack of optimization flag results in kernels (from cuBINs) using up to 4 additional registers in 62% of the total kernels. However, when we use full optimizations in the compilation (O3), 71% of kernels use no extra registers, 13% use up to one extra register, and 7% use up to two extra registers. In some rare cases, the number of registers is smaller than the default because the compiler spills some registers in the global memory.

### 7.4 Performance of Fencing at High Cache Hit Ratio

Figure 10 shows the overhead of Guardian address fencing (bitwise operation) normalized to native for 890000 kernels used in lenet. The overhead of Guardian is, on average 3.2%. We have performed the same breakdown for computer vision and observed similar results. The overheads of Guardian bit-masking instructions depend on the latency of the load and store instructions. A load instruction that retrieves data from global memory is 220-350 cycles [7, 25], while if data are in L1-cache is 28 cycles. Our approach adds two (bitwise AND, OR) up to four instructions (for cases that include address+offset) per load and store instruction. Each of these instructions is executed in almost 4 cycles. As a result, if all data are in the L1 cache –not common–, our overhead is from 28% up to 57%. If all data are only in global memory, our overhead (with global memory latency 285 cycles) is from 2% up to 5%. We have profiled all kernels of lenet and observed that the average L1 cache hit rate is 37%, while for L2 is 72%. L2 latency is 180 cycles, only 1.4× better than global's. Overall, address fencing (bitwise operation) in Guardian incurs small additional overhead due to two main reasons: (1) As we show, ML kernels exhibit a low cache hit ratio. (2) As shown from previous works [4], cache hits result in a lower load/store instruction latency in the rear case that every thread in the warp hits in the cache.

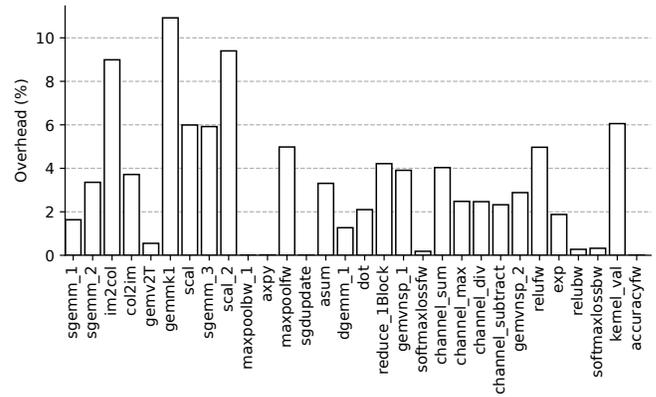

Figure 10: Performance overhead of sandboxed kernels against native execution.

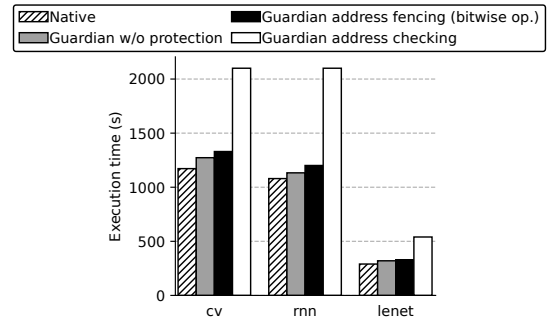

Figure 11: Guardian overhead with PyTorch and Caffe on GeForce GPU, compared to native execution.

### 7.5 Evaluation on Other GPUs and Access Patterns

Figure 11 shows neural networks from PyTorch and Caffe executed in the GeForce GPU. In computer vision (cv) and rnn Guardian address fencing (bitwise operation) incurs 12% and 10% overhead compared to native, respectively. Lenet with Guardian incurs 13% overhead compared to native. Conditional checks exhibit on average 1.8× worst execution time compared to native. Overall, we note that Guardian has similar overhead across different GPU types.

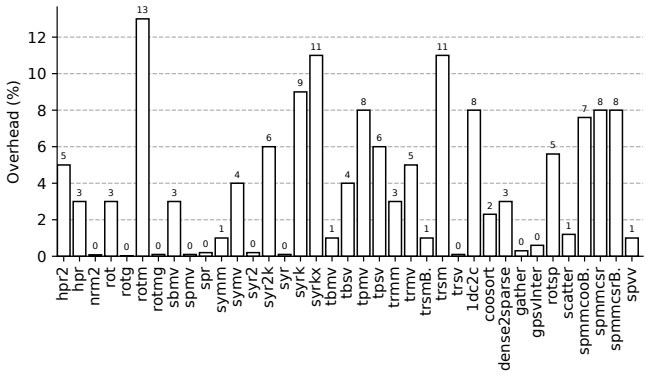

Figure 12: Guardian overhead (%) for 37 kernels from CUDA-accelerated libraries compared to native execution of each kernel on the GeForce GPU.

|  | Lookup GPU kernel | Augment kernel params | Launch kernel to GPU |
|---|---|---|---|
| Native | 0 | 0 | ~9000 |
| Guardian | 557 | 400 | ~9000 |

Table 5: Guardian average cost in CPU cycles for the main operations performed when a kernel launch is intercepted and replaced with a sandboxed kernel.

| High-level call | Implicit CUDA runtime/driver calls | Total |
|---|---|---|
| cublasCreate | cudaMalloc: 3<br>cudaEventCreateWithFlags: 18<br>cudaFree: 2 | 23 |
| cublasIdamax | cudaLaunchKernel: 1<br>cudaMemcpy: 1<br>cudaEventRecord:1<br>cudaStreamGetCaptureInfo:2 | 5 |
| cublasDdot | cudaLaunchKernel: 2<br>cudaMemcpy: 1<br>cudaEventRecord: 1<br>cudaStreamGetCaptureInfo:2 | 6 |
| cusparseAxpby | cudaLaunchKernel: 2 | 2 |
| cufftExecC2C | cuMemcpyHtoD: 2<br>cuMemAlloc: 1<br>cuMemFree: 1<br>cuLaunchKernel: 1<br>cudaStreamIsCapturing: 1 | 6 |
| cusolverSpDcsrqr | cudaLaunchKernel: 2<br>cuMemcpyHtoD: 1<br>cuMemAlloc: 1 | 4 |

Table 6: Implicit CUDA runtime and drive calls performed from high-level function calls of CUDA accelerated libraries.

Figure 12 shows the performance of Guardian over CUDA-accelerated library calls that are not contained in the ML frameworks used previously. Guardian successfully intercepts these calls and adds 4% overhead, on average, which is similar to the results observed with the Quadro GPU.

### 7.6 Cost of CUDA calls Interception

The interception of kernel invocations in Guardian requires between 214 and 900 CPU cycles ("Lookup GPU kernel" in Table 5) for the lookup operation to locate the sandboxed kernel (stored in a c++ unordered map named *pointerToSymbol*). Regarding the extra arguments passed in the kernel, we require between 300 and 600 CPU cycles to allocate a new parameter array and copy the new and old parameters in this array ("Augment kernel params" in Table 5). Guardian adds on average 957 CPU cycles per cudaLaunchKernel. We perform each experiment ten times, excluding the minimum and maximum values.

The cudaLaunchKernel NVIDIA system call is measured using the Nsight profiler. The average execution time (for more than one thousand kernels) in CPU cycles is approximately 9000 CPU cycles ("Launch kernel to GPU" in Table 5). So our overhead *without the kernel execution* is 10% on average. We have profiled lenet and cv applications (executing millions of kernels) and found out that the kernel execution time *without the cudaLaunchKernel* is, on average 18000 CPU cycles. Consequently, the overhead of Guardian, including the kernel execution time, is 3% per kernel, on average.

*Memory allocation and data transfer.* We use a micro-benchmark that uses memory allocations and data transfers of different sizes to evaluate Guardian's memory management operations. The results suggest that (a) our allocator does not imply overhead compared to native CUDA, and (b) the protection checks used on every data transfer over the PCIe bus imply negligible overhead.

### 7.7 Implicit calls from Closed-source Libraries

Table 6 shows the implicit calls to CUDA runtime and driver API performed from high-level function calls of three libraries namely cuBLAS, cuSPARSE, and cuFFT. As we observe, they perform tens of CUDA calls to both CUDA runtime and driver APIs. Consequently, treating these calls as black boxes is inadequate for Guardian since they could lead to illegal memory accesses.

## 8 RELATED WORK

This section reviews related work on GPU memory protection in shared environments, buffer overflow detection without sharing, privacy-preserving and data confidentiality. It also covers techniques for intercepting high-level function calls and improving GPU sharing efficiency, highlighting key advancements in security and performance.

**Protecting GPU memory under GPU sharing.** Mask [6] is a hardware-based approach that allows applications to share spatially and securely a GPU. Mask extends the GPU TLB to hold information about warps and the memory they can use, and as a result, it supports protected spatial-sharing. Guardian does not need extra or special hardware and implies comparable overhead, making it more practical, powerful, and generic.

G-NET [82] is a software-based approach that deploys a custom type of pointer [57], which checks if a pointer belongs to the correct partition. However, leveraging these pointers requires manual effort to port the kernel source code. The source code requirement is a serious limitation leading to weaker protection because most CUDA-accelerated applications rely heavily on closed-source GPU libraries [64]. Guardian operates in the PTX level of kernel code available in closed-source accelerated libraries.

**Detecting buffer overflows without sharing.** clArmor [16] and GMOD [13] protect against overflows by adding canary values around the allocated buffers. However, such approaches have limited security coverage because they cannot capture non-adjacent accesses that jump over canaries. Parravicini et al. [50] add conditional checks inside the kernel LLVM-IR to preserve Java memory safety semantics in NVIDIA GPUs. They use static analysis to minimize the significant overhead implied by conditional checks, which require the application and kernel source code (or LLVM-IR), limiting its applicability. GPUShield [30] overcomes the limitation of source code using an extra hardware unit that performs the address checking. CUDA-MEMCHECK and cuCatch [60] are debugging tools that operate in the PTX [45] level and detect out-of-bounds accesses without requiring extra/specific hardware. All these approaches focus on buffer overflow detection of a single application and are considered orthogonal to Guardian.

**Preserving privacy and data confidentiality.** Graviton [65] is a trusted execution environment (TEE) providing privacy and data confidentiality guarantees. Graviton requires minimal hardware modifications only in the GPU command processor. Honeycomb [36] relies on address checking to eliminate the necessity for hardware modifications. Furthermore, it depends on source code for static analysis to minimize its overhead. NVIDIA H100 offers a TEE [2] for GPUs that in cooperation with MIG allows spatial-sharing. Although, TEEs tackle a significantly different problem [46], Guardian can be combined with them to provide trusted execution under spatial-sharing but with low overhead.

**Intercepting high-level function calls.** Cricket [15], DGSF [18], rCUDA [14], Arax [52], and GPUless [61] treats high-level functions to CUDA accelerated libraries as a black box. This is because CUDA libraries use an undocumented function, the cudaGetExportTable(), which exports a set of function pointers that implement hidden functionalities. Guardian intercepts only the CUDA runtime and driver API that are performed implicitly by the high-level functions of CUDA accelerated libraries. To achieve this, Guardian provides a minimal implementation of the cudaGetExportTable() hidden CUDA calls, which is adequate to run PyTorch and Caffe. We experiment with both interception approaches and determine that the Guardian interception approach is more robust. This is because we only need to intercept 400 relatively straightforward CUDA runtime-driver API calls, as opposed to dealing with 1600 high-level (far more complex) calls to CUDA-accelerated libraries [52].

**Improving GPU sharing.** Several approaches, including Zico [35], Planaria [20], Pagoda [78], Orion [58], Wavelet [66], GPUlet [10], DGSF [18] utilize multiple streams and a single context to enable spatial-sharing. Different to Guardian these approaches primarily focus on scheduling rather than memory isolation. Guardian can complement these techniques by integrating memory isolation features while benefiting from their advanced scheduling policies.

## 9 CONCLUSIONS

This paper introduces Guardian, a mechanism for isolating GPU faults from memory accesses, making GPU sharing among various real-world applications both practical and efficient. The benefits of Guardian are threefold: (1) It is transparent to applications, even when applications use closed-source GPU-accelerated libraries that include host and device code. (2) It fences all memory accesses (even from closed-source GPU kernels) using instrumentation at the PTX level. (3) It uses bit-masking instructions to optimize address fencing. Our evaluation on real-world ML frameworks shows that Guardian can support ML frameworks and CUDA-accelerated libraries transparently, introducing 9% overhead (average) compared to native unprotected execution.

## ACKNOWLEDGMENTS

We thank our shepherd Fabio Kon for his help preparing the final version of the paper and the anonymous reviewers for their insightful comments. We thankfully acknowledge the support of EUPILOT (GA No 101034126). European PILOT has received funding from the European High-Performance Computing Joint Undertaking (EuroHPC JU) under grant agreement No 101034126. The JU receives support from the European Union's Horizon 2020 research and innovation programme and Spain, Italy, Switzerland, Germany, France, Greece, Sweden, Croatia, and Turkey.